\documentclass[12pt]{article}
\usepackage[cp1251]{inputenc}
\usepackage[russian]{babel}
\begin{document}
\large
\begin{center}
{\bf Schemes of Neutrino Mixings (Oscillations) and Their Mixing
Matrices}
\par
\vspace{0.5cm} Beshtoev Kh. M.
\par
\vspace{0.5cm} Joint Institute for Nuclear Research., Joliot Curie
6, 141980 Dubna, Moscow region and Institute Applied Mathematics
and Automation  KBSC of RAS, Nalchik,  Russia; \\
\end{center}

\par
{\bf Abstract}
\par
Three schemes of neutrino mixings (oscillations) together with
their mixing matrices (analogous to Kabibbo-Kobayashi-Maskawa
matrices) are considered.  In these schemes neutrino transitions
are virtual if neutrino masses are different. Two of them belong
to the so called mass mixing schemes (mixing parameters are
expressed by elements of mass matrices) and the third scheme
belongs to the charge mixing scheme (mixing parameters are
expressed through charges). In the first scheme system of 6
equations for determination of the all elements of the mass matrix
(neutrino masses and transition widths) by using experimental data
are obtained. In the second and third ones the neutrino mixing
angles are equal or close to maximal angles ($\pi/4$). It is
obvious that the experiment must give an answer to the following
question: Which of these schemes is realized indeed? \\

\section{Introduction}

\par
In the quark sector, mixings between $d, s, b$ quarks are
described by Kabibbo-Kobayashi-Maskawa matrices [1]. At present,
we know that the lepton numbers are not conserved [2-5] and
$\nu_e, \nu_\mu, \nu_\tau$ neutrinos are also mixed. Then, for
lepton sector we can also introduce similar matrices.
Unfortunately, we do not know: are there neutrino oscillations, or
only neutrino mixings without oscillations take place? Therefore,
it is necessary to consider all the realistic schemes of neutrino
mixings and oscillations. Usually, only the standard scheme of
neutrino oscillations is considered [6]. Since in this scheme the
law of energy-momentum conservation is not fulfilled [7], we
suppose that this scheme is not a realistic one for description of
neutrino oscillations. We proposed three schemes for description
of neutrino mixings and oscillations [7]. The first scheme is the
development of the standard scheme in the framework of the
particle physics. In these schemes neutrino transitions are
virtual if neutrino masses are different. You are invited to study
these schemes of neutrino mixings and oscillations. In this paper
we also obtain mixing matrices for these schemes.

\section{Schemes (Types) of Neutrino Mixing (Oscillation) and Their Mixing Matrices}

\par
In a common case there can be two schemes (types) of neutrino
mixings (oscillations): mass mixing schemes and charge  mixings
scheme (as it takes place in the vector dominance model or vector
boson mixings in the standard model of electroweak
interactions).\\

\subsection{Two Schemes of Neutrino Mass Mixings (Oscillations) and Their Mixing Matrices}

\par
In the standard approach [6] it is supposed that neutrinos are
once created in superposition states, i.e., mass  matrix is a
nondiagonal one. If  mass matrix is nondiagonal at once, then we
must diagonalize this matrix in order to find eigenstates of
neutrinos. Then eigenstates are $\nu_1, \nu_2, \nu_3$ neutrinos,
i.e., there must be created $\nu_1, \nu_2, \nu_3$ neutrinos but
not $\nu_e , \nu_\mu, \nu_\tau$ neutrinos. It is obvious that it
cannot be coordinated with experimental data. In the weak
interactions only physical neutrinos ($\nu_e , \nu_\mu, \nu_\tau$)
are created, i.e., at once mass matrix is a diagonal one, and then
at violation of the lepton numbers this matrix is transformed into
nondiagonal one [7]. We stress this point for its fundamental
importance.
\par
Originally it was supposed [6] that these neutrino oscillations
are real oscillations, i.e., that there takes place a real
transition of electron neutrino $\nu_e$ into muon neutrino
$\nu_{\mu}$ (or tau neutrino $\nu_{\tau}$). Then the neutrino $x =
\mu, \tau$ will decay in electron neutrino plus something
$$
\nu_{x} \rightarrow \nu_e + ....  , \eqno(1)
$$
as a result, we get energy from vacuum, which equals the mass
difference (if $m_{\nu_x} > m_{\nu_e}$)
$$
\Delta E \sim m_{\nu_{x}} - m_{\nu_e} . \eqno(2)
$$
Then, again this electron neutrino transits into muon  neutrino,
which decay again and we get energy and etc. {\bf So we got a
perpetuum mobile!} Obviously, the law of energy conservation
cannot be fulfilled in this process. The only way to restore the
law of energy conservation is to demand that this process is a
virtual one. Then, these oscillations will be the virtual ones and
they are described in the framework of the uncertainty relations.
The correct theory of neutrino oscillations can be constructed
only  in the framework of the particle physics theory, where the
concept of mass shell is present [8], [9].
\par
We can also see that there are two cases of neutrino
transitions (oscillations) in the scheme of mass mixings [9]. \\

\par
\subsubsection{Development of the Standard Scheme of Neutrino
Mixings (Oscillations)}

\par
The standard scheme belongs to the so-called mass mixings scheme,
since mixing parameters are expressed through elements of mass
matrix. In this case the probability of $\nu_e \rightarrow
\nu_\mu$ transition (oscillation) is described by the following
expression (for simplification we consider two neutrino $\nu_e,
\nu_\mu$ mixings cases):
$$
P(\nu_e \rightarrow \nu_\mu, t) =  sin^2 2\theta sin^2 \left[\pi
t\frac{\mid m_{\nu_1}^2 - m_{\nu_2}^2 \mid}{2 p_{\nu_e}} \right ],
\eqno(3)
$$
where $p_{\nu_e}$ is a momentum of $\nu_e$ neutrino,
$$
sin^2 2\theta = \frac{4m^2_{\nu_e, \nu_\mu}}{(m_{\nu_e} -
m_{\nu_\mu})^2 + 4m^2_{\nu_e, \nu_\mu}}  , \eqno(4)
$$
and
$$
m_{1, 2} = {1\over 2} \left[ (m_{\nu_e} + m_{\nu_\mu}) \pm
\left((m_{\nu_e} - m_{\nu_\mu})^2 + 4 m^{2}_{\nu_\mu \nu_e}
\right)^{1/2} \right] , \eqno(5)
$$
At this transitions (oscillations) neutrinos remain on their mass
shell and transitions (oscillations) must be virtual.
\par
It is interesting to remark that expression (4) can be obtained
from the Breit-Wigner distribution [10]
$$
P \sim \frac{(\Gamma/2)^2}{(E - E_0)^2 + (\Gamma/2)^2}   ,
\eqno(6)
$$
by using the following substitutions:
$$
E = m_{\nu_e},\hspace{0.2cm} E_0 = m_{\nu_\mu},\hspace{0.2cm}
\Gamma/2 = 2m_{\nu_e, \nu_\mu} , \eqno(7)
$$
where $\Gamma/2 \equiv W(... )$ is a width of $\nu_e \rightarrow
\nu_\mu$ transition, then we can use a standard method [9, 11] for
calculating this value. Then, the probability of $\nu_e
\rightarrow \nu_\mu$ transitions is defined by these neutrino
masses and width of their transitions.
\par
Expression for length of these oscillations has the following
form:
$$
L_{o} = 2\pi  {2p \over {\mid m^{2}_{2} - m^{2}_{1} \mid}} .
\eqno(8)
$$
\par
Above, we considered the case of two neutrino transitions
(oscillations). In the common case we must consider three neutrino
transitions (oscillations). For a complete description of three
neutrino oscillations we must have six parameters (we suppose that
this mass matrix is symmetric in respect of the diagonal one),
$$
\left(\begin{array}{ccc}m_{\nu_e} & m_{\nu_e \nu_\mu} & m_{\nu_e \nu_\tau} \\
m_{\nu_e \nu_\mu} & m_{\nu_\mu} & m_{\nu_\mu \nu_\tau}
\\ m_{\nu_e \nu_\tau} &  m_{\nu_\mu \nu_\tau} & m_{\nu_\tau} \end{array} \right) ,
\eqno(9)
$$
three diagonal terms of this matrix are masses of three physical
neutrinos $m_{\nu_\mu}, m_{\nu_\tau}, m_{\nu_e}$ and three
nondiagonal mass terms of this matrix are $m_{\nu_e \nu_\mu}$,
$m_{\nu_\mu \nu_\tau}, m_{\nu_e \nu_\tau}$-neutrino transition
widthes. Since in the expression for neutrino transition
probabilities mass differences (in squared form) are used in
reality, we need only five parameters (and for further
simplification the physical neutrino masses are used). Besides, if
mass matrix is complex, there appears one parameter, connected
with $CP$ violation.
\par
Let us consider back problem, i.e., problem of finding of these
(six) parameters from experiments. From experiments on neutrino
transitions (oscillations) we can determine the following six
values:
\par
\noindent
three values from amplitudes
$$
sin^2 2\theta_{ij} = \frac{4m^2_{\nu_i \nu_j}}{(m_{\nu_i} -
m_{\nu_j})^2 + 4m^2_{\nu_i \nu_j}}, \eqno(10)
$$
and three values from oscillation lengths (or differences of
squared masses $m_{\nu_\alpha}^2 - m_{\nu_\beta}^2$)
$$
P'(\nu_i \rightarrow \nu_j, t) =  sin^2 \left[\pi t \frac{\mid
m_{\nu_\alpha}^2 - m_{\nu_\beta}^2 \mid}{2 p_{\nu_i}} \right ],
\eqno(11)
$$
where
$$
i < j, \qquad i, j = e, \mu, \tau; \qquad \alpha, \beta = 1, 2, 3
$$
Using these parameters we can obtain values of six neutrino mass
matrix parameters: three values for neutrino mass (or two mass
differences) and three nondiagonal mass parameters (widths of
neutrino transitions).
\par
These mixing angles can be connected with the mixing matrix $V$ in
the same manner as it takes place for the
Kabibbo-Kobayashi-Maskawa matrices [1]. We will choose a
parameterization of  the mixing matrix $V$ in the form proposed by
Maiani [12]:
\par
$$
{V = \left( \begin{array} {ccc}1& 0 & 0 \\
0 & c_{\gamma} & s_{\gamma} \\ 0 & -s_{\gamma} & c_{\gamma} \\
\end{array} \right) \left( \begin{array}{ccc} c_{\beta} & 0 &
s_{\beta} \exp(-i\delta) \\ 0 & 1 & 0 \\ -s_{\beta} \exp(i\delta)
& 0 & c_{\beta} \end{array} \right) \left( \begin{array}{ccc}
c_{\theta} & s_{\theta} & 0 \\ -s_{\theta} & c_{\theta} & 0 \\ 0 &
0 & 1 \end{array}\right)} , \eqno(12)
$$
\par
$$
c_{e \mu} = \cos {\theta } , \quad s_{e \mu} =\sin{\theta}, \quad
c^2_{e \mu} + s^2_{e \mu} = 1 ;
$$
$$
c_{e \tau} = \cos {\beta }, \quad s_{e \tau} =\sin{\beta}, \quad
c^2_{e \tau} + s^2_{e \tau} = 1 ; \eqno(13)
$$
$$
c_{\mu \tau} = \cos {\gamma} , \quad s_{\mu \tau} =\sin{\gamma},
\quad c^2_{\mu \tau} + s^2_{\mu \tau} = 1 ;
$$
$$
 \exp(i\delta) = \cos{\delta } + i \sin{\delta} .
$$
In our approximation, the value of $\delta$ can be considered to
be equal to zero.
\par
Equations for mixing angles expressed through elements of mass
matrix have the following form:
$$
s_{e \mu} = \sin {\theta } = \frac{1}{\sqrt{2}} \left[ 1 -
\frac{\mid m_{\nu_\mu} - m_{\nu_e} \mid}{\sqrt{(m_{\nu_\mu} -
m_{\nu_e})^2 + (2 m_{\nu_e \nu_\mu})^2}} \right] , \eqno(14)
$$
\par
$c^2_{e \mu} = 1 - s^2_{e \mu}$ ;
$$
s_{e \tau} = \sin {\beta } = \frac{1}{\sqrt{2}} \left[ 1 -
\frac{\mid m_{\nu_\tau} - m_{\nu_e} \mid}{\sqrt{(m_{\nu_\tau} -
m_{\nu_e})^2 + (2 m_{\nu_e \nu_\tau})^2}} \right] , \eqno(15)
$$
\par
$c^2_{e \tau} = 1 - s^2_{e \tau}$ ;
$$
s_{\mu \tau} = \sin {\gamma} = \frac{1}{\sqrt{2}} \left[ 1 -
\frac{\mid m_{\nu_\tau} - m_{\nu_\mu} \mid}{\sqrt{(m_{\nu_\tau} -
m_{\nu_\mu})^2 + (2 m_{\nu_\mu \nu_\tau})^2}} \right] , \eqno(16)
$$
\par
$c^2_{\mu \tau} = 1 - s^2_{\mu \tau}$ .\\

\subsubsection{Analysis of Present Status of Neutrino Mixing
Parameters}

\par
Super-Kamiokande data [3] on atmospheric neutrino transitions for
$\nu_\mu \to \nu_\tau$ are
$$
sin^2 2 \beta \cong 1, \quad \Delta m^2_{2 3} \cong 2.5\cdot
10^{-3} eV^2 . \eqno(17)
$$
The KamLAND detector [5] on $\bar \nu_e \to \bar \nu_\mu$
transitions presented the following data:
$$
sin^2 2\theta \cong 1, \quad \Delta m^2_{2 1} \cong 6.9 \cdot
10^{-5} eV^2 . \eqno(18)
$$
Using the above and the SNO [4] data we can come to conclusion
that for $\nu_e \to \nu_\tau$ transitions we have
$$
sin^2 2 \gamma \cong 1, \eqno(19)
$$
but the value of $\Delta m^2_{3 1}$ remains not fixed.
\par
The vicinity of $sin^2 2 \theta, sin^2 2\beta, sin^2 2\gamma$ to
unity allows possibility for the expansion of these values around
unity and then expressions for $sin^2 2\theta_{i j}$ and mass
differences will have the following form:
\par
$ (2 m_{ij})^2 \gg (m_j - m_i)^2 \quad j>i;   i,j = e, \mu, \tau$
$$
sin^2 2\theta_{ij} \cong 1 - \frac{(m_{\nu_i} -
m_{\nu_j})^2}{4m^2_{\nu_i, \nu_j}}, \eqno(20)
$$
$$
\Delta m^2_{2 1} = m^2_2 - m^2_1 = (m_{\nu_\mu} + m_{\nu_e})
\sqrt{(m_{\nu_\mu} - m_{\nu_e})^2 + (2 m_{\nu_e \nu_\mu})^2},
\eqno(21)
$$
if $ 2 m_{\nu_e \nu_\mu} \gg \mid m_{\nu_\mu} - m_{\nu_e} \mid $
then
$$
\Delta m^2_{2 1} = (m_{\nu_\mu} + m_{\nu_e}) 2 m_{\nu_e \nu_\mu}
\left[1 + \frac{(m_{\nu_\mu} - m_{\nu_e})^2}{2(2 m_{\nu_e
\nu_\mu})^2}\right] ,\eqno(21')
$$
and if $ 2 m_{\nu_e \nu_\mu} \ll \mid m_{\nu_\mu} - m_{\nu_e} \mid
$ then
$$
\Delta m^2_{2 1} = (m^2_{\nu_\mu} - m^2_{\nu_e}) \left[1 +
\frac{(2 m_{\nu_e \nu_\mu})^2} {2(m_{\nu_\mu} -
m_{\nu_e})^2}\right] ; \eqno(21'')
$$
$$
\Delta m^2_{3 1} = m^2_3 - m^2_1 = (m_{\nu_\tau} + m_{\nu_e})
\sqrt{(m_{\nu_\tau} - m_{\nu_e})^2 + (2 m_{\nu_e \nu_\tau})^2},
\eqno(22)
$$
if  $2 m_{\nu_e \nu_\tau} \gg \mid m_{\nu_\tau} - m_{\nu_e} \mid $
then
$$
\Delta m^2_{3 1} = (m_{\nu_\tau} + m_{\nu_e})2 m_{\nu_e \nu_\tau}
\left[1 + \frac{(m_{\nu_\tau} - m_{\nu_e})^2}{2 (2 m_{\nu_e
\nu_\tau})^2} \right] , \eqno(22')
$$
and if  $2 m_{\nu_e \nu_\tau} \ll \mid m_{\nu_\tau} - m_{\nu_e}
\mid $ then
$$
\Delta m^2_{3 1} = (m^2_{\nu_\tau} - m^2_{\nu_e}) \left[1 + \frac{
(2 m_{\nu_e \nu_\tau})^2}{2(m_{\nu_\tau} - m_{\nu_e})^2} \right] ;
\eqno(22'')
$$
$$
\Delta m^2_{3 1} = m^2_3 - m^2_2 = (m_{\nu_\tau} + m_{\nu_\mu})
\sqrt{(m_{\nu_\tau} - m_{\nu_\mu})^2 + (2 m_{\nu_\mu
\nu_\tau})^2}, \eqno(23)
$$
if $2 m_{\nu_\mu \nu_\tau} \gg \mid m_{\nu_\tau} - m_{\nu_\mu}
\mid $ then
$$
\Delta m^2_{3 1} = (m_{\nu_\tau} + m_{\nu_\mu})2 m_{\nu_\mu
\nu_\tau} \left[1 + \frac{(m_{\nu_\tau} - m_{\nu_\mu})^2}{2 (2
m_{\nu_\mu \nu_\tau})^2} \right] , \eqno(23')
$$
and if $2 m_{\nu_\mu \nu_\tau} \ll \mid m_{\nu_\tau} - m_{\nu_\mu}
\mid $ then
$$
\Delta m^2_{3 1} = (m^2_{\nu_\tau} - m^2_{\nu_\mu}) \left[1 +
\frac{(2 m_{\nu_\mu \nu_\tau})^2}{2(m_{\nu_\tau} - m_{\nu_mu})^2}
\right] . \eqno(23'')
$$
\par
In common case from the neutrino oscillation experiments, we can
obtain six values $ sin^2 2\theta, sin^2 2\beta, sin^2 2\gamma$,
$\Delta m_{21}, \Delta m_{32}, \Delta m_{31}$, which   can be used
for determination of six parameters $m_{\nu_e}, m_{\nu_\mu},
m_{\nu_\tau}$, $m_{\nu_e \nu_\mu}, m_{\nu_\mu \nu_\tau}, m_{\nu_e
\nu_\tau}$ using 3 equations (20) for $sin^2 2 ...$ and  3
equations (21)-(23). Unfortunately, these equation are
transcendental ones  and they can be solved only numerically.
\par
In the simplest case
$$
sin^2 2\theta \cong sin^2 2\beta \cong sin^2 2\gamma \cong 1,
\eqno(24)
$$
$$
 m_{\nu_e} \cong m_{\nu_\mu} \cong m_{\nu_\tau} = m_\nu
 \eqno(25)
$$
and we get
$$
\Delta m^2_{21} \cong 4m_{\nu_e \nu_\mu} \cdot m_{\nu} \cong 2.5
\cdot 10^{-3} eV^2 ,
$$
$$
\Delta m^2_{31} \cong 4m_{\nu_e \nu_\tau}\cdot m_{\nu} \cong
\hspace{0.15cm} while \hspace{0.15cm} remains \hspace{0.15cm}
unknown , \eqno(26)
$$
$$
\Delta m^2_{32} \cong 4m_{\nu_e \nu_\mu} \cdot m_{\nu} \cong 6.9
\cdot 10^{-5} eV^2 ,
$$
and there is no possibility to obtain values of masses
of physical neutrinos.\\

\subsubsection{The Case of Neutrino Mixings without Mass Shell
Changing}

 \par
Above we considered the case when virtual neutrino transitions
take place with change of neutrino masses. Another case is also
possible, when $\nu_e$ neutrino transits into $\nu_\mu$ neutrino
without changing mass, i. e., $m^{*}_{\nu_\mu} = m_{\nu_e}$ then
$$
tg 2\theta = \infty  , \eqno(27)
$$
$\theta = \pi/4$, and
$$
sin^2 2\theta = 1     . \eqno(28)
$$
\par
In this case the probability of the $\nu_e \rightarrow \nu_\mu$
transition (oscillation) is described by the following expression:
$$
P(\nu_e \rightarrow \nu_\mu, t) = sin^2 \left[\pi t\frac{4
m_{\nu_e, \nu_\mu}^2}{2 p_a} \right ] . \eqno(29)
$$
\par
Expression for length of oscillations this case has the following
form:
$$
L_{o} = 2\pi  \frac{2p} {(2m_{\nu_e \nu_\mu})^2}  .
$$
\par
In order to make these virtual oscillations real, their
participation in quasi-elastic interactions is necessary for their
transitions to their own mass shells [9].
\par
The matrix, analogous to Kabibbo-Kobayashi-Maskawa in this case,
is a trivial one and it has the following form:
$$
{V = \left( \begin{array} {ccc}1& 0 & 0 \\
0 & c_{\gamma} & s_{\gamma} \\ 0 & -s_{\gamma} & c_{\gamma} \\
\end{array} \right) \left( \begin{array}{ccc} c_{\beta} & 0 &
s_{\beta} \exp(-i\delta) \\ 0 & 1 & 0 \\ -s_{\beta} \exp(i\delta)
& 0 & c_{\beta} \end{array} \right) \left( \begin{array}{ccc}
c_{\theta} & s_{\theta} & 0 \\ -s_{\theta} & c_{\theta} & 0 \\ 0 &
0 & 1 \end{array}\right)} , \eqno(30)
$$
\par
$$
c_{e \mu} = \cos {\theta } = \frac{1}{\sqrt2} , \quad s_{e \mu}
=\sin{\theta}= \frac{1}{\sqrt2} ;
$$
$$
c_{e \tau} = \cos {\beta }= \frac{1}{\sqrt2}, \quad s_{e \tau}
=\sin{\beta}= \frac{1}{\sqrt2} ; \eqno(31)
$$
$$
c_{\mu \tau} = \cos {\gamma}= \frac{1}{\sqrt2} , \quad s_{\mu
\tau} =\sin{\gamma}= \frac{1}{\sqrt2} ;
$$
$$
 \exp(i\delta) = 1 .
$$
In our approximation the value of $\delta$ can be considered to be
equal to zero.
\par
In the case
$$
sin^2 2\theta = sin^2 2\beta = sin^2 2\gamma = 1, \eqno(32)
$$
we have
$$
\Delta m^2_{21} = (2m_{\nu_e \nu_\mu})^2 ,
$$
$$
\Delta m^2_{31} = (2m_{\nu_e \nu_\tau})^2 , \eqno(33)
$$
$$
\Delta m^2_{32} = (2m_{\nu_e \nu_\mu})^2 ,
$$
and we can obtain values of nondiagonal mass terms (widths of
neutrino transitions) but there is no possibility of obtaining
values of masses of physical neutrinos.
\par
It is necessary to remark that in physics all the processes are
realized through dynamics. Unfortunately, in this mass mixings
scheme the dynamics is absent. Probably, that is  an indication of
the fact that these schemes are incomplete ones, i.e., these
schemes demand a physical substantiation (see section 2.2).
\par
Obviously, these schemes will only work if neutrino oscillations
take place in reality (it is clear that there also can be
neutrino mixings in absence of neutrino oscillations).\\

\par
\subsection{The Scheme of Neutrino Charge Mixings (Oscillations)}

\par
The third scheme (type) of neutrino mixings or transitions can be
realized by mixings of the neutrino fields by analogy with the
vector dominance model ($\gamma-\rho^o$ and $Z^o-\gamma$ mixings)
in the same way as it takes place in the particle physics. Then,
in the case of two neutrinos, we have
$$
\nu_1 = cos \theta \nu_e - sin \theta \nu_\mu , \eqno(34)
$$
$$
\nu_2 = sin \theta \nu_e + cos \theta \nu_\mu .
$$
In the case of three neutrinos we can also choose parameterization
of the mixing matrix $V$ in the form proposed by Maiani [12]:
$$
{V = \left( \begin{array} {ccc}1& 0 & 0 \\
0 & c_{\gamma} & s_{\gamma} \\ 0 & -s_{\gamma} & c_{\gamma} \\
\end{array} \right) \left( \begin{array}{ccc} c_{\beta} & 0 &
s_{\beta} \\ 0 & 1 & 0 \\ -s_{\beta} & 0 & c_{\beta}
\end{array} \right) \left( \begin{array}{ccc} c_{\theta} &
s_{\theta} & 0 \\ -s_{\theta} & c_{\theta} & 0 \\ 0 & 0 & 1
\end{array}\right)} , \eqno(35)
$$
$$
c_{e \mu} = \cos {\theta } \quad s_{e \mu} =\sin{\theta}  ,
 \quad c^2_{e \mu} + s^2_{e \mu} = 1 ;
$$
$$
c_{e \tau} = \cos {\beta }, \quad s_{e \tau} =\sin{\beta} , \quad
c^2_{e \tau} + s^2_{e \tau} = 1 ; \eqno(36)
$$
$$
c_{\mu \tau} = \cos {\gamma} , \quad s_{\mu \tau} =\sin{\gamma} ,
\quad c^2_{\mu \tau} + s^2_{\mu \tau} = 1 ;
$$

The charged current in the standard model of weak interactions for
two lepton families has the following form:
$$
j^\alpha  = \left(\begin{array}{cc} \bar e \bar \mu
\end{array}\right)_L \gamma^\alpha V \left(\begin{array}{c} \nu_e \\
\nu_\mu \end{array} \right)_L ,
$$
$$
V = \left(\begin{array}{cc} \cos \theta & \sin \theta \\
-\sin \theta & cos \theta \end{array}\right) , \eqno(37)
$$
and then the interaction Lagrangian is
$$
{\cal L} = \frac{g}{\sqrt{2}} j^\alpha W^{+}_\alpha  + h.c.
\eqno(38)
$$
and
$$
\begin{array}{c}
\nu_e = cos \theta  \nu_{1} + sin \theta \nu_{2}           \\
\nu _{\mu } = - sin \theta  \nu_{1} + cos \theta  \nu_{2} .
\end{array}
\eqno(39)
$$
Then, taking into account that the charges of $\nu_1, \nu_2$
neutrinos are $g_1, g_2$ we have
$$
g cos \theta = g_1, \quad g sin \theta = g_2 , \eqno(40)
$$
i. e.,
$$
cos \theta = \frac{g_1}{g}, \quad sin \theta = \frac{g_2}{g}.
\eqno(41)
$$
Since $sin^2 \theta + cos^2 \theta = 1$, then
$$
g = \sqrt{g_1^2 + g_2^2}
$$
and
$$
cos \theta = \frac{g_1}{\sqrt{g_1^2 + g_2^2}}, \quad sin \theta =
\frac{g_2}{\sqrt{g_1^2 + g_2^2}}. \eqno(42)
$$
\par
Since we suppose that $ g_1 \cong g_2 \cong \frac{g}{\sqrt{2}}$,
then
$$
cos \theta \cong  sin \theta \cong \frac{1}{\sqrt{2}} . \eqno(43)
$$
It is not difficult to come to consideration of the case of three
neutrino types $\nu_e, \nu_{\mu}, \nu_\tau$. Since the weak couple
constants $g_{\nu_e}, g_{\nu_{\mu}}, g_{\nu_{\tau}}$ of $\nu_e,
\nu_{\mu}, \nu_{\tau}$  neutrinos are approximately equal in
reality, i.e., $g_{\nu_e}\simeq g_{\nu_{\mu}} \simeq
g_{\nu_{\tau}}$ then the angle mixings are nearly maximal:
$$
 cos \theta = cos \theta_{\nu_e \nu_\mu} \cong  sin \theta_{\nu_e \nu_\mu} \cong
\frac{1}{\sqrt{2}} ,
$$
$$
cos \beta = cos \theta_{\nu_e \nu_\tau} \cong  sin \theta_{\nu_e
\nu_\tau} \cong \frac{1}{\sqrt{2}} , \eqno(44)
$$
$$
cos \gamma = cos \theta_{\nu_\mu \nu_\tau} \cong  sin
\theta_{\nu_\mu \nu_\tau} \cong \frac{1}{\sqrt{2}} .
$$
\par
As it is stressed above in the case of mass mixings scheme, we
have no dynamical substantiation in contrast to the case of charge
mixings scheme, but these schemes may be jointed if neutrino
masses have the following form:
$$
m_{\nu_i} = g_i v, \quad i = e, \mu, \tau, \eqno(45)
$$
where $v$ is constant, as it is in the Higgs mechanism [13]. And
then the problem of dynamical substantiation in this scheme is
solved.\\

\section{Conclusion}

\par
Unfortunately, we do not know there are neutrino oscillations or
only neutrino mixings without oscillations. Therefore, it is
necessary to consider all the realistic schemes of neutrino
mixings and oscillations. In this work three schemes of neutrino
mixings (oscillations) together with their mixing matrices
(analogous to Kabibbo-Kobayashi-Maskawa matrices) were considered
.  In these schemes neutrino transitions are virtual if neutrino
masses are different. Two of them belong to the so-called mass
mixing schemes (mixing parameters are expressed by elements of
mass matrices) and the third scheme belongs to the charge mixings
one (mixing parameters are expressed through charges). For the
first scheme, the equations for determination of the all elements
of mass matrix (neutrino masses and transition widths) by using
experimental data were given. In the second and third ones the
mixing angles are equal or close to the maximal angles ($\pi/4$).
It is obvious that the experiment must get an answer to
the following question: Which of these schemes is realized indeed? \\

\par
{\bf References}

\par
\noindent 1. Cabibbo N., Phys. Rev. Lett., 1963, v.10, p.531;
\par
Kobayashi M., Maskawa K., Prog. Theor. Phys., 1973, v.49, p.652.
\par
Review of part. Prop., Phys. Rev.D, 1994, v.50, N 2.
\par
\noindent 2. Kameda J., Proceedings of ICRC 2001, August 2001,
Germany,
\par
Hamburg, p.1057.
\par
Fukuda  S. et al,. Phys.   Rev . Lett. 2001, v. 25, p.5651;
\par
Phys. Lett. B 539, 2002,  p.179.
\par
\noindent 3. Toshito T., hep-ex/0105023;
\par
Kameda J., Proceeding of 27th ICRC, August 2001;
\par
Hamburg, Germany, v.2, p.1057.
\par
Mauger Ch., 31-st ICHEP, Amsterdam, July 2002.
\par
\noindent 4. Ahmad Q. R. et al., Internet Pub. nucl-ex/0106015,
June 2001.
\par
Ahmad  Q. R. et al., Phys. Rev. Lett. 2002, v. 89, p.011301-1;
\par
Phys. Rev. Lett.  2002,v.  89, p.011302-1.
\par
\par
\noindent 5. Mitsui T., Proceedings of the 31-st ICHEP, August
2002,
\par
Amsterdam, Netherlands.
\par
Eguchi K. et al., Phys. Rev. Lett. 2003, v.90, p.021802.
\par
\noindent 6. Bilenky S. M., Pontecorvo B. M., Phys. Rep. C, 1978,
v.41, p.225;
\par
Boehm F., Vogel P., Physics of Massive Neutrinos: Cambridge
\par
Univ. Press, 1987;
\par
Bilenky S. M., Petcov S. T., Rev. of Mod.  Phys., 1977, v.59,
\par
p.631.
\par
\noindent 7. Beshtoev Kh.M., JINR Commun., E2-2003-155, Dubna,
2003;
\par
 Proceed. of 28th International Cosmic Ray
Conference, Japan,
\par
2003, V.1, p.1503;.
\par
Proceed. of 28th International Cosmic Ray Conference, Japan,
\par
2003, V.1, p.1507.
\par
\noindent 8. Beshtoev Kh.M., JINR Commun. E2-92-318, Dubna, 1992;
\par
JINR Rapid Communications, N3[71]-95.
\par
\noindent 9. Beshtoev Kh.M., Internet Pub. hep-ph/9911513;
\par
 The Hadronic Journal, v.23, 2000, p.477;
\par
Proceedings of 27th Intern. Cosmic Ray Conf., Germany,
\par
Hamburg, 7-15 August 2001, v.3, p. 1186.
\par
\noindent 10. Blatt J.M., Waiscopff V.F., The Theory of Nuclear
Reactions,
\par
INR T.R. 42.
\par
\noindent 11. Beshtoev Kh.M., JINR Commun. E2-99-307, Dubna, 1999;
\par
JINR Commun. E2-99-306, Dubna, 1999.
\par
\noindent 12. L. Maiani, Proc. Inter. Symp.  on  Lepton-Photon
\par
Inter., Hamburg, DESY, p.867.
\par
\noindent 13. Higgs P.W., Phys. Lett., 1964, v.12, p.132;
Phys.Rev., 1966,
\par
v.145, p.1156;
\par
Englert F., Brout R., Phys. Rev. Lett, 1964, v.13, p.321;
\par
Guralnik G.S., Hagen C.R.,  Kible  T.W.B., Phys.  Rew.  Lett,
1964,
\par
vol.13, p.585.

\end{document}